\documentclass[preprint,12pt,numbers]{elsarticle}

\usepackage{graphicx}
\usepackage{booktabs}
\usepackage{array}
\usepackage{enumitem}
\usepackage{xcolor}
\usepackage{listings}
\usepackage{caption}
\usepackage{amsmath,amssymb}
\usepackage{longtable}
\usepackage{tabularx}
\usepackage{microtype}
\usepackage[hidelinks]{hyperref}
\newenvironment{revision}{}{}

\journal{Journal of Systems and Software}
\biboptions{numbers,sort&compress}

\newcolumntype{Y}{>{\raggedright\arraybackslash}X}
\newcolumntype{L}[1]{>{\raggedright\arraybackslash}p{#1}}

\lstdefinestyle{code}{
  basicstyle=\ttfamily\footnotesize,
  breaklines=true,
  frame=single,
  framerule=0.3pt,
  backgroundcolor=\color{gray!5},
  columns=fullflexible,
  keepspaces=true,
  showstringspaces=false
}

\lstdefinelanguage{yaml}{
  keywords={true,false,null,y,n},
  keywordstyle=\color{black},
  basicstyle=\ttfamily\footnotesize,
  sensitive=false,
  comment=[l]{\#},
  morecomment=[l]{//},
  morestring=[b]',
  morestring=[b]"
}

\begin{document}

\begin{frontmatter}

\title{From Conventional Multi-Vendor Failover to Adaptive API Routing: An Industrial Experience Report on Resilient Third-Party Service Integration}

\author[aff1]{Nataraj Agaram Sundar\corref{cor1}}
\ead{natsundar@ebay.com}
\author[aff1]{Tejas Morabia}
\address[aff1]{eBay Inc., San Jose, CA, United States}
\cortext[cor1]{Corresponding author.}

\begin{abstract}
High-scale online services often rely on third-party APIs in user-facing flows such as authentication, messaging, payments, fraud detection, and identity verification. Integrating alternate providers and operating conventional failover controls is a common resilience baseline, but redundancy alone does not make provider selection adaptive, explainable, or policy-aware. This paper reports an anonymized industrial experience evolving a conventional multi-vendor SMS-provider failover arrangement in a large marketplace setting into configuration-driven adaptive API routing. The approach uses operation-specific pluggable factor lists to separate routing policy from application code, combines hard eligibility gates with normalized weighted provider scoring, and closes the loop with business-outcome telemetry, decision logs, traffic-shift controls, and recovery safeguards. The report includes de-identified operational evidence, including before/after windows, a representative incident timeline, shadow-mode validation records, policy-tuning history, score-normalization snapshots, provider-attribution examples, feedback-bias controls, and sensitivity analysis. The evidence suggests that adaptive provider routing can reduce dependence on incident-time interpretation when the operation is critical, telemetry volume is sufficient, and organizational controls exist for policy ownership, explainability, and safe traffic movement. The paper concludes with practitioner guidance and transferability conditions for teams considering similar architectures.
\end{abstract}


\end{frontmatter}

\section{Introduction}

Third-party APIs are often part of critical user journeys. A login flow can depend on identity verification, SMS delivery, email delivery, risk evaluation, device reputation, and internal profile services. A checkout flow can depend on tax, payments, fraud detection, inventory, shipping, and notification providers. The user experiences the combined behavior of these dependencies, not the individual availability number of any one provider. When an external dependency degrades, the owning application team cannot patch the provider, roll back its deployment, or directly repair its infrastructure. These concerns are consistent with prior work on large-scale service failures, service reliability engineering, and the operational complexity of microservice-based systems \cite{oppenheimer2003internet,beyer2016sre,newman2021microservices,diFrancesco2019architecting,wang2021promises}.

A common response is to integrate more than one provider and operate those providers through a primary/secondary failover model, health checks, dashboards, alerts, configuration controls, and incident runbooks. This is a reasonable and common industry baseline; conventional controls such as timeouts, retries, circuit breakers, service-mesh traffic policies, and runbooks remain necessary resilience mechanisms \cite{nygard2018release,sedghpour2021servicemesh,sedghpour2022empirical}. The issue was that the conventional arrangement did not continuously reason about business outcomes, region-specific eligibility, quota, completion rate, latency, cost, confidence, and recovery behavior as part of a governed request-time decision. When provider behavior degraded partially or regionally, nuanced traffic movement still depended on incident-time diagnosis, operator judgment, and configuration changes.

This paper is an industrial experience report on evolving that conventional multi-vendor failover model into configuration-driven adaptive API routing. The target contribution is not a new load-balancing algorithm, a new circuit-breaker implementation, or a generic service-mesh mechanism. Instead, we report how known reliability mechanisms were combined in an industrial service-integration context to address a practical problem: making third-party provider selection more adaptive, explainable, policy-aware, and less dependent on incident-time operator judgment. The work is related to self-adaptive microservice research and feedback-control approaches, but it is deliberately positioned as a practitioner-facing account of a governed runtime decision layer in a third-party API integration setting \cite{hellerstein2004feedback,mendonca2021selfadaptive,figueira2024developing,jawabreh2026selfadaptation}.

The architecture uses pluggable factor lists: operation-specific configuration objects that define which providers are eligible and how eligible providers should be ranked. Eligibility is controlled by hard gates such as region support, compliance allowance, quota availability, credential validity, and circuit-breaker state. Preference among eligible providers is controlled by weighted scores such as recent completion rate, p95 latency, cost, incident penalty, and sample-size confidence. A telemetry pipeline converts business outcomes into rolling provider-health metrics. Decision logs record the policy version, eligible providers, gate outcomes, scores, selected provider, and reason code. Safety controls such as traffic-shift caps, cooldowns, shadow mode, and kill switches reduce the risk that automation creates a new incident.

The paper is organized around what a practitioner or researcher can learn from the experience. Section~\ref{sec:context} describes the industrial setting and limits of the conventional failover model. Section~\ref{sec:why-not-enough} explains why familiar mechanisms were insufficient by themselves. Section~\ref{sec:rationale} presents the design rationale and architecture. Section~\ref{sec:rollout} describes implementation and rollout. Section~\ref{sec:case-study} summarizes the SMS verification case. Section~\ref{sec:challenges} discusses challenges encountered and mitigations. Section~\ref{sec:evaluation} presents a supporting synthetic evaluation. Section~\ref{sec:lessons} distills lessons learned and practitioner guidance. Section~\ref{sec:transferability} discusses transferability, limitations, and threats to validity.

\paragraph{Contributions}
This paper contributes an industrial experience report on the design, deployment, and operation of a configuration-driven API routing mechanism for resilient third-party service integration. The report explains how operation-specific factor lists, runtime policy separation, hard eligibility gates, weighted scoring, business-event telemetry, and traffic-shift safety controls were combined to extend a conventional vendor-failover baseline with adaptive routing in an SMS verification setting. The contribution is not a claim that conventional failover is inadequate in general; rather, it identifies the conditions under which a conventional baseline benefits from an adaptive decision layer. It further contributes a discussion of industrial context, design rationale, adoption challenges, transferability conditions, and practitioner lessons for teams considering similar architectures.

\paragraph{Intended reader takeaways}
The report is written for practitioners and researchers who need to decide whether adaptive routing is worth its operational cost. The main takeaways are: (1) a multi-provider integration becomes more resilient when provider choice is treated as a governed runtime decision, not only as a static priority; (2) business-outcome telemetry and decision explainability are as important as the scoring function; (3) safety controls such as shadow mode, traffic caps, cooldowns, and kill switches are part of the architecture rather than optional operational add-ons; and (4) the approach should be adopted only where criticality, traffic volume, provider substitutability, and governance maturity justify the added complexity.

\paragraph{Relationship to prior work}
The experience complements prior work on microservice architecture, service reliability engineering, service-mesh traffic management, self-adaptive systems, and chaos engineering \cite{jamshidi2018journey,chen2018microservices,sedghpour2022empirical,owotogbe2026chaos}. Those streams explain many of the underlying mechanisms and evaluation concerns. The distinction in this report is the industrial combination of these ideas for third-party provider selection, where the decision must account for provider eligibility, business outcome telemetry, policy governance, and operator trust rather than only balancing traffic among homogeneous internal service instances.

\section{Industrial Context and Limits of Conventional Failover}
\label{sec:context}

The experience described here comes from a large marketplace environment in which user-facing workflows depend on third-party services. We anonymize provider names, internal system names, exact production metrics, and proprietary thresholds. The relevant operation in this report is SMS verification for login and registration. SMS verification is operationally important because a failure can prevent legitimate users from signing in or completing account creation. It is also difficult to reason about because provider behavior can vary by country, carrier, message type, compliance constraint, throttling policy, and time window.

Before the adaptive-routing work, the platform already had more than one SMS provider integrated. This was the expected industry baseline rather than an absence of resilience engineering. Provider selection followed a conventional primary/secondary model supported by monitoring, alerts, operational configuration, and incident procedures. In degradation windows, the conventional process typically involved the following activities:

\begin{enumerate}[leftmargin=*]
  \item An on-call engineer or operations dashboard detected abnormal login or registration verification behavior.
  \item The team investigated whether the issue was caused by the application, a regional carrier condition, a third-party provider, a configuration change, or downstream verification logic.
  \item If the issue appeared provider-specific, the team checked whether the secondary provider was contractually, technically, and operationally eligible for the affected traffic.
  \item The team changed routing configuration or executed a runbook to move traffic.
  \item Engineers monitored completion rate, latency, retry behavior, and user-facing error indicators.
  \item After the primary provider appeared healthy, the team decided when to move traffic back.
\end{enumerate}

The practical problem was not that no failover existed. The problem was that the conventional model was not sufficiently adaptive for partial, regional, quota-related, or business-outcome degradation. During a provider incident, engineering attention should be focused on diagnosis, containment, communication, and recovery. When nuanced provider selection depends on incident-time interpretation, teams must reason under pressure about eligibility, health, regional behavior, quota, and recovery. That creates delay, inconsistent decisions, and weaker post-incident analysis when the reason for each routing decision is not captured in a structured, queryable form.

\subsection{Events leading to the adaptive-routing work}

The adaptive-routing work was motivated by a recurring operational pattern rather than by the absence of backup infrastructure. The platform had conventional resilience controls, but provider degradation could still appear as an ambiguous operational signal: verification completion would decline in a region or carrier group, transport-level errors might not fully explain the decline, and the healthiest routing action depended on eligibility, quota, cost, recovery state, and confidence in the telemetry. In those situations, the team needed a mechanism that could evaluate the same decision dimensions consistently for every request and still leave operators with visibility and override authority.

Table~\ref{tab:baseline-evolution} summarizes the change in control model. The adaptive layer did not remove existing protections; it added a policy-driven decision layer above them.

\begin{table}[htbp]
\centering
\caption{Evolution from conventional failover to adaptive routing in the reported setting.}
\label{tab:baseline-evolution}
\small
\begin{tabularx}{\linewidth}{L{0.24\linewidth}YY}
\toprule
Decision dimension & Conventional multi-provider failover & Adaptive routing enhancement \\
\midrule
Provider eligibility & Primary/secondary posture with operational checks. & Request-time gates for region, compliance, quota, credentials, circuit state, and metric freshness. \\
Health signal & Provider-level dashboards and alerts. & Operation- and region-aware rolling metrics from business outcome events. \\
Traffic movement & Configuration change or runbook action after diagnosis. & Controlled shifts subject to traffic caps, cooldowns, and reason codes. \\
Recovery & Operator judgment supported by monitoring. & Probe traffic, half-open behavior, cooldowns, and gradual failback. \\
Explainability & Incident notes and dashboard snapshots. & Decision logs with policy version, gate outcomes, score components, selected provider, and outcome. \\
Governance & Operational ownership of failover procedure. & Versioned factor-list ownership across engineering, SRE, compliance, finance, and product concerns. \\
\bottomrule
\end{tabularx}
\end{table}

\subsection{Operational constraints}

Several constraints shaped the design.

\begin{itemize}[leftmargin=*]
  \item \textbf{Regional variability.} A provider could be healthy globally but degraded in a specific country, carrier group, or operation type.
  \item \textbf{Business outcome versus transport success.} A successful HTTP response from an SMS provider did not necessarily mean that the verification flow succeeded. Completion-rate telemetry had to be tied to business outcomes where possible.
  \item \textbf{Policy constraints.} Region support, contractual limits, compliance requirements, credential state, and quota availability were not optimization preferences; they were eligibility requirements.
  \item \textbf{Incident safety.} Moving too much traffic too quickly could create a secondary incident, exhaust quota, or mask recovery signals from the original provider.
  \item \textbf{Operational trust.} On-call engineers needed to understand why the router moved traffic. A black-box routing decision would not be acceptable during an incident.
  \item \textbf{Confidentiality.} Production logs and exact vendor performance metrics could not be shared externally. This influenced how evidence is presented in this paper and why a synthetic artifact accompanies it.
\end{itemize}

\begin{table}[htbp]
\centering
\caption{Industrial context that motivated adaptive routing.}
\label{tab:context}
\small
\begin{tabularx}{\linewidth}{L{0.27\linewidth}Y}
\toprule
Context element & Practical implication for the routing design \\
\midrule
Critical user journey & Login and registration verification could not rely solely on a conventional primary/secondary failover model and incident runbook. \\
Multiple external providers & Redundancy existed, but provider choice still required policy, health, and safety decisions. \\
Regional and carrier variation & Routing decisions had to be operation-specific and region-aware, not only provider-wide. \\
Business-outcome telemetry & Completion behavior mattered more than transport-level availability alone. \\
Confidential industrial setting & Public evaluation had to use anonymized context and synthetic replay artifacts. \\
\bottomrule
\end{tabularx}
\end{table}

\section{Why Conventional Mechanisms Were Not Enough}
\label{sec:why-not-enough}

The solution did not discard established resilience patterns. Timeouts, retries, circuit breakers, bulkheads, dashboards, and runbooks remained necessary. The practical issue was that each mechanism solved only part of the problem. Table~\ref{tab:mechanisms} summarizes what each mechanism contributed and where it fell short in this setting.

\begin{table}[htbp]
\centering
\caption{Why familiar mechanisms were useful but insufficient alone.}
\label{tab:mechanisms}
\small
\begin{tabularx}{\linewidth}{L{0.22\linewidth}Y Y}
\toprule
Mechanism & What it helped with & Why it was insufficient alone \\
\midrule
Timeouts and retries & Avoided indefinite waits and handled transient failures. & Retrying a degraded provider could amplify load, increase latency, and still leave provider selection unchanged. \\
Circuit breakers & Protected local resources and stopped traffic to a dependency that crossed failure thresholds. & Circuit state was often binary and local. It did not rank multiple eligible providers using region, completion rate, quota, cost, or policy context. \\
Bulkheads & Isolated thread pools, connection pools, and worker capacity. & Bulkheads contained failure but did not decide where future requests should go. \\
Static primary/secondary routing & Provided a common and useful fallback baseline. & It could not continuously adapt to partial degradation, regional variation, quota exhaustion, or provider recovery using business-outcome telemetry. \\
Incident runbooks and operational overrides & Preserved human judgment and explicit accountability. & They were necessary safety tools, but they were not a substitute for routine adaptive decisioning when degradation was partial, regional, or fast-moving. \\
Generic load balancing & Distributed traffic across endpoints. & The decision needed to incorporate business outcome, compliance, region eligibility, incident penalty, and recovery dampening. \\
\bottomrule
\end{tabularx}
\end{table}

\begin{revision}
\subsection{Boundary between established practices and the contribution}
\label{sec:contribution-boundary}

To clarify the boundary between established engineering practices and the contribution reported here, Table~\ref{tab:contribution-boundary} makes this distinction explicit. The contribution is the industrial composition and governance of established mechanisms into a closed-loop provider-routing control plane; it is not a claim that circuit breakers, service meshes, load balancing, shadow mode, or observability are new in isolation.

\begin{table}[htbp]
\centering
\caption{Established practices versus the adaptive-routing extension reported here.}
\label{tab:contribution-boundary}
\scriptsize
\begin{tabularx}{\linewidth}{L{0.20\linewidth}YY}
\toprule
Component & Established practice & Adaptive-routing extension in this experience \\
\midrule
Circuit breaker & Common distributed-systems resilience pattern. & Used as a non-negotiable eligibility gate before scoring, with half-open probes and velocity-capped failback. \\
Business outcome telemetry & Standard observability and monitoring practice. & Used as real-time closed-loop input: downstream verification completion updates provider completion-rate indices. \\
Shadow mode & Standard canary and verification technique. & Used for continuous parallel evaluation of third-party routing policies against live traffic before enablement. \\
Traffic shifting & Load balancing and disaster-recovery failover. & Extended with exploration, hysteresis score margins, and step-velocity caps to avoid oscillation during partial outages. \\
\bottomrule
\end{tabularx}
\end{table}
\end{revision}

This analysis drove a design principle: local protection mechanisms should remain in the request path, but provider selection should be treated as a separate control-plane decision. The control plane needed to encode both non-negotiable eligibility rules and changeable ranking preferences. It also needed to produce explanations for decisions so that operators could trust, override, and audit it.

\section{Design Rationale and Architecture}
\label{sec:rationale}

\begin{figure}[htbp]
  \centering
  \includegraphics[width=0.95\linewidth]{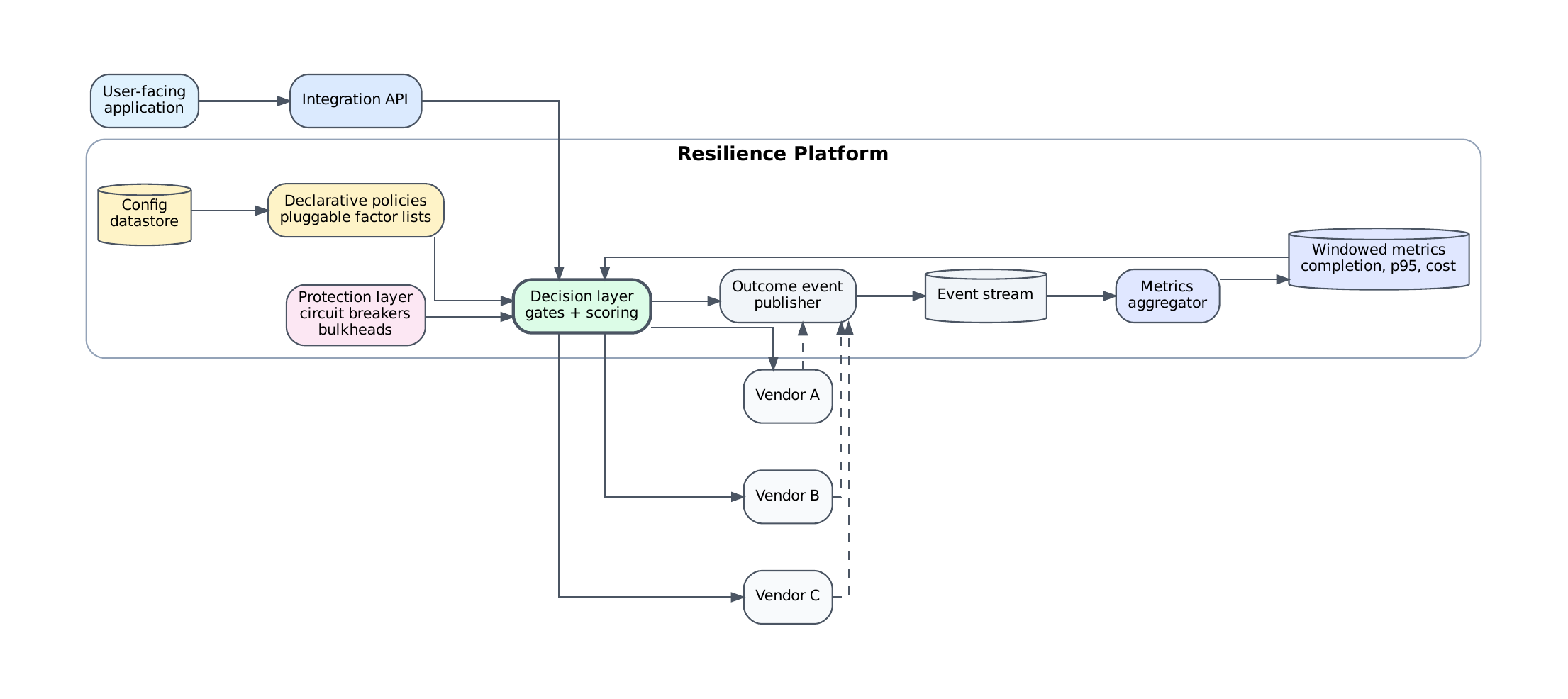}
  \caption{Layered architecture for adaptive third-party provider routing. The architecture combines local protection, a decision layer, event telemetry, and configuration-controlled routing policy.}
  \label{fig:architecture}
\end{figure}

Figure~\ref{fig:architecture} shows the architecture. The platform separates four concerns. The protection layer contains timeouts, retries, circuit breakers, bulkheads, and fallback behavior. The decision layer selects a provider for an operation. The telemetry layer publishes the outcome of each provider call and computes rolling metrics. The configuration control plane stores factor lists and safety settings.

The design was shaped by practical questions and boundary cases that repeatedly surfaced during design and rollout.

\subsection{Why separate routing policy from application code?}

Application code should express the business operation, such as sending a verification message. It should not contain all rules for ranking providers by completion rate, latency, quota, cost, region, policy, and incident state. Those rules change more frequently than core application behavior and often require input from SRE, product, compliance, finance, and vendor-management stakeholders. Moving routing policy into versioned configuration allowed changes to be reviewed, rolled back, and audited without redeploying the request-path service.

\subsection{Why use hard gates before scoring?}

Some constraints should not be traded off. A provider that is not approved for a region should not win because it is cheap. A provider with an open circuit should not win because it has good historical latency. The factor-list model therefore applies hard gates before weighted scoring. Gates encode eligibility; scores encode preference. This separation made routing decisions easier to explain during incidents and reduced the risk of optimization overriding safety.

\subsection{Why use business-event telemetry?}

Transport-level metrics were necessary but incomplete. For SMS verification, a provider API call could succeed while the user still failed to complete verification. Conversely, a provider could return an accepted response but later experience carrier-level delivery issues. The routing system therefore treated provider-call outcomes as business events that included operation type, provider, region, status, latency, error class, retry count, decision reason, and policy version. Where available, downstream verification completion events were used to compute completion-rate signals.

\subsection{Why use cached rolling metrics?}

A synchronous dependency on an analytics pipeline in the request path would have introduced another availability risk. The design used an asynchronous event pipeline and cached rolling-window metrics. This reduced request-path coupling but introduced a new challenge: stale metrics. The factor list therefore included metric freshness checks, sample-size confidence, and safe fallback behavior when telemetry was delayed.

\subsection{Why include decision logs?}

Automated routing can erode trust if operators cannot explain it. Each decision logged the operation, request region, policy version, candidate providers, gate outcomes, normalized scores, selected provider, reason code, and outcome. During incidents, this allowed the team to answer questions such as: Which gate excluded the primary provider? Did cost weighting influence the decision? Was a metric stale? Did a policy change precede the traffic shift?

\subsection{Why add dampening and recovery controls?}

A naive scoring system can oscillate. If traffic shifts away from a provider, the system observes fewer fresh samples for that provider. If it shifts back too quickly, it can re-trigger the failure. The design used traffic-shift caps, cooldown windows, half-open circuit behavior, and probe traffic. These mechanisms made recovery gradual and observable rather than abrupt.

\subsection{Factor-list example}

Listing~\ref{lst:factor} shows a simplified factor list for SMS routing. The exact production configuration is not disclosed, but the example captures the structure used in the experience: gates first, scores second, and safety controls around traffic movement.

\begin{lstlisting}[style=code, language=yaml, caption={Simplified factor list for SMS routing.}, label={lst:factor}]
operation: SEND_SMS
refresh_interval_seconds: 30
metric_window_minutes: 5
min_sample_size: 200
fallback_policy: use_secondary_then_queue

gates:
  - supports_region
  - compliance_allowed
  - quota_available
  - circuit_breaker_closed
  - metric_fresh_or_safe_default

scores:
  completion_rate_5m:        {weight: 0.45, direction: higher}
  p95_latency_5m:            {weight: 0.20, direction: lower}
  cost_per_request:          {weight: 0.15, direction: lower}
  recent_incident_penalty:   {weight: 0.10, direction: lower}
  confidence_from_sample:    {weight: 0.10, direction: higher}

safety:
  max_traffic_shift_per_minute: 0.50
  primary_probe_share: 0.01
  primary_recovery_cooldown_seconds: 300
  require_reason_code: true
\end{lstlisting}

\begin{revision}
\subsection{Score normalization, weighting, and policy evaluation}
\label{sec:score-normalization}

The operational evidence extract provides decision-level and snapshot-level evidence for how heterogeneous metrics were combined. The router first applied hard gates. Only providers passing those gates were scored. Raw metrics were converted to bounded 0--100 index values so that completion rate, p95 latency, cost index, incident penalty, and sample-size confidence could be combined without mixing incompatible units. A representative policy version used:
\[
\begin{aligned}
score(v) ={}& 0.40 \cdot completion(v) + 0.25 \cdot latency(v) \\
&+ 0.15 \cdot cost(v) + 0.10 \cdot incident(v) \\
&+ 0.10 \cdot confidence(v).
\end{aligned}
\]
The weights reflected an operational priority order: completion quality first, latency second, cost third, and incident/confidence safeguards as stabilizers. They were not learned by an optimizer. They were initially selected by expert review, exercised in shadow mode, then refined through canary rollout and post-incident review. For example, the tuning history shows a latency-weight increase from 0.15 to 0.25 after canary validation showed p95 degradation sensitivity, and a later hysteresis threshold change from 2.0 to 3.5 after post-incident review identified oscillation risk.

\begin{table}[htbp]
\centering
\caption{Example normalized scoring snapshot from the operational evidence extract.}
\label{tab:score-snapshot}
\scriptsize
\begin{tabular}{lrrrrrrr}
\toprule
Provider & Completion & Latency & Cost & Incident & Confidence & Final score & Rank \\
\midrule
provider\_A & 104.0 & 105.0 & 95.0 & 100.0 & 100.0 & 96.5 & 1 \\
provider\_B & 98.0 & 92.0 & 100.0 & 100.0 & 98.0 & 88.2 & 2 \\
provider\_C & 90.0 & 80.0 & 85.0 & 100.0 & 88.0 & 81.4 & 3 \\
\bottomrule
\end{tabular}
\end{table}

This design keeps the score interpretable. A provider can be ranked first because its normalized completion and latency evidence outweighs cost differences, but it cannot bypass an eligibility gate. The policy version, factor-list version, score components, score margin, selected reason code, metric age, and staleness flag are logged with each decision so that operators can inspect the decision after the fact.

\subsection{Provider-specific attribution of business outcomes}
\label{sec:provider-attribution}

Verification completion may be influenced by user behavior, downstream application logic, or provider behavior. The evidence model therefore separates provider-call outcomes from business-completion outcomes. Provider-level events record latency, HTTP/status buckets, error class, timeout, throttle, quota, regional error, circuit state, and provider-attributable failure flags. Business-completion events then record whether the user completed verification within the SLA window and whether the failure stage was before the provider call, at the provider call, during user action, in downstream application logic, or unknown.

This separation does not make completion rate a perfect causal estimate of provider quality. Instead, it makes it a safer operational routing signal. Provider-attributable failures are isolated using request/decision correlation, error classification, retry stage, and failure-stage labeling. User abandonment, invalid input, and downstream application errors are separated from provider-attributable degradation before the provider metric is used for routing.

\subsection{Feedback-bias controls during traffic shifts}
\label{sec:feedback-bias}

Adaptive routing changes the data it observes. When traffic shifts away from a degraded provider, fewer production samples are collected from that provider, making recovery harder to detect. The evidence extract shows three safeguards: a minimum probe share, a minimum sample-size requirement, and staged recovery through cooldown and half-open states. During the representative degradation window, provider A retained a 3\% probe share while its production traffic share was reduced. Probe samples kept the actual sample size above the configured minimum of 100, allowing recovery to be detected before full failback.

\begin{table}[htbp]
\centering
\caption{Feedback-bias controls in the representative incident window.}
\label{tab:probe-controls}
\scriptsize
\begin{tabular}{lrrrrrr}
\toprule
Time & Traffic share & Probe share & Sample size & Min sample & Score after probe & Recovered? \\
\midrule
2026-06-14 09:20 & 15.0\% & 3.0\% & 725 & 100 & 61.8 & false \\
2026-06-14 09:25 & 15.0\% & 3.0\% & 742 & 100 & 78.4 & true \\
2026-06-14 09:30 & 25.0\% & 3.0\% & 1228 & 100 & 94.2 & true \\
\bottomrule
\end{tabular}
\end{table}
\end{revision}

\begin{revision}
\subsection{Boundary cases and conflicting routing objectives}
\label{sec:boundary-cases}

Three boundary cases are important in practice.

First, if all available providers are simultaneously degraded, the router should not blindly select a provider merely because it is ranked first. The factor-list model distinguishes hard eligibility gates from preference scoring. If every provider fails a hard gate, such as compliance allowance, credential validity, region support, quota availability, or circuit state, the routing layer invokes an explicit fallback policy rather than forcing a provider selection. Examples include queueing the request for later retry, presenting degraded user messaging, rate-limiting or throttling non-critical requests, routing only emergency probe traffic, or escalating to operator-controlled provider pinning. If multiple providers remain eligible but all are below a soft service objective, the router may choose the least-bad provider only within predefined guardrails, traffic caps, and escalation rules.

Second, there is an intentional trade-off between reacting quickly to genuine degradation and avoiding unnecessary traffic shifts caused by short-lived noise. In the reported design, hard signals such as open circuit state, quota exhaustion, unsupported region, or invalid credentials can trigger immediate exclusion. Softer signals such as completion-rate decline or p95 latency increase are evaluated using rolling windows, minimum sample-size thresholds, metric freshness checks, hysteresis, cooldown windows, and probe traffic. This means the system can react quickly to unambiguous failure while requiring more evidence before shifting traffic for noisy or transient changes. The appropriate balance is operation-specific: login verification may justify faster protective movement than a lower-criticality notification flow.

Third, routing objectives can disagree. A provider may have the highest completion rate but worse latency or higher cost. The design handles this in layers. Hard gates first remove providers that are not allowed or safe to use. Soft guardrails then define unacceptable operating regions, such as latency above a service objective, cost above a budget threshold, or insufficient sample confidence. Weighted scoring is used only after those constraints are satisfied. Therefore, the weighted score is not the only control mechanism; it is the final preference mechanism among candidates that remain operationally acceptable. In high-criticality flows, completion and safety should dominate; in lower-criticality flows, cost and latency weights may be more influential.

\begin{table}[htbp]
\centering
\caption{Boundary cases and routing behavior.}
\label{tab:boundary-cases}
\scriptsize
\begin{tabularx}{\linewidth}{L{0.25\linewidth}YY}
\toprule
Boundary case & Routing behavior & Safety control \\
\midrule
All providers fail hard gates & Do not force provider selection; invoke fallback policy. & Queue, retry later, degraded message, operator escalation, or probe-only routing. \\
All providers are degraded but eligible & Select the least-bad provider only within guardrails. & Traffic caps, probe traffic, escalation threshold, and operator override. \\
Short-lived metric fluctuation & Avoid immediate full traffic shift. & Rolling windows, minimum samples, hysteresis, cooldowns, and freshness checks. \\
Hard failure signal & Exclude the provider immediately. & Circuit gate, quota gate, region gate, compliance gate, and credential gate. \\
Conflicting objectives & Apply gates and soft guardrails before weighted scoring. & SLO threshold, cost cap, confidence threshold, and policy-specific weight profile. \\
High completion but high latency or cost & Use only if the provider remains within operational limits. & Latency/cost guardrails, traffic caps, and escalation rules. \\
\bottomrule
\end{tabularx}
\end{table}
\end{revision}

\section{Implementation and Rollout}
\label{sec:rollout}

The change was not enabled globally in one step. The rollout followed an incremental path designed to build operational confidence.

\begin{table}[htbp]
\centering
\caption{Rollout stages used to reduce adoption risk.}
\label{tab:rollout}
\small
\begin{tabularx}{\linewidth}{L{0.22\linewidth}Y Y}
\toprule
Stage & What was done & Why it mattered \\
\midrule
Instrumentation & Provider calls emitted business events with operation, region, provider, latency, status, decision metadata, and policy version. & Reliable telemetry was a prerequisite for safe automated decisions. \\
Dashboarding & Rolling completion rate, latency, error class, circuit state, and traffic share became visible by provider and region. & Operators needed to see both provider health and router behavior. \\
Shadow mode & The decision layer computed the provider it would have selected while production traffic continued on existing routing. & Teams could compare recommendations with observed incidents without moving traffic. \\
Limited enablement & Dynamic routing was enabled for a narrow operation/region combination before broader rollout. & The blast radius of a bad routing decision was limited. \\
Operational controls & Kill switches, provider pinning, cooldowns, traffic caps, and rollback procedures were documented. & Automation remained reversible during incidents. \\
Post-incident review & Decision logs were reviewed alongside provider metrics and user-facing indicators. & Lessons from incidents fed back into factor weights, gates, and dashboards. \\
\bottomrule
\end{tabularx}
\end{table}

The shadow-mode stage was especially important. It helped answer whether the proposed factor list would have reacted to known degradation windows, whether it would have moved traffic unnecessarily during normal variance, and whether the chosen metrics were fresh enough. Shadow mode also exposed disagreements about what counted as success. For example, provider-level API success, message delivery indicators, and user verification completion are related but not identical. The team had to decide which signals should be used for routing and which should remain diagnostic.

Governance was another practical concern. A factor list is configuration, but configuration can materially change user traffic. The operating model treated routing policy changes with similar discipline to production code changes. Ownership had to be clear: service teams owned operation-specific factor lists, SRE owned default safety controls, compliance stakeholders owned regional gates, and operational leaders owned incident override procedures.

\begin{revision}
\subsection{Shadow-mode validation}
\label{sec:shadow-validation}

The operational evidence extract clarifies how shadow-mode recommendations were validated before production enablement. In shadow mode, the adaptive decision engine evaluated live request contexts and produced the provider it would have selected, while the actual routing path continued to use the conventional baseline. The validation checked that shadow recommendations passed gates, aligned with known degradation windows, did not recommend providers with open circuits or blocked quotas, and were reviewed by operators before enablement.

\begin{table}[htbp]
\centering
\caption{Representative shadow-mode validation records.}
\label{tab:shadow-validation}
\scriptsize
\begin{tabular}{llllll}
\toprule
Time & Actual & Shadow rec. & Actual outcome & Incident? & Avoided degraded? \\
\midrule
T-15d & provider\_A & provider\_A & SUCCESS & false & false \\
T-12d\_inc & provider\_A & provider\_B & FAILURE\_5XX & true & true \\
T-12d\_inc & provider\_A & provider\_B & FAILURE\_TIMEOUT & true & true \\
\bottomrule
\end{tabular}
\end{table}

The representative shadow evidence shows one stable case where the shadow policy matched the conventional baseline and two incident cases where the shadow policy recommended provider B while the conventional baseline continued to route to degraded provider A. These records are not presented as a complete production experiment; they are included to explain the validation method and the type of evidence used before enabling adaptive routing.
\end{revision}

\begin{revision}
\subsection{Policy tuning and governance}
\label{sec:policy-tuning}

The routing policy was tuned through a hybrid process rather than a purely automated optimizer. Initial weights came from expert judgment about the relative operational importance of completion rate, latency, cost, incident penalties, and confidence. Shadow-mode and canary evidence then informed adjustments, and post-incident reviews produced targeted parameter updates. Table~\ref{tab:policy-tuning} summarizes the policy-version evidence available in the anonymized evidence extract.

\begin{table}[htbp]
\centering
\caption{Policy-version and tuning history from the anonymized evidence extract.}
\label{tab:policy-tuning}
\scriptsize
\resizebox{\linewidth}{!}{%
\begin{tabular}{llllll}
\toprule
Policy & Change type & Factor & Source & Validation & Rollback \\
\midrule
v1.0.0-shadow & initial\_deployment & all\_weights & initial\_design & expert\_review & v0.0.0-legacy \\
v1.1.0-prod & weight\_rebalance & latency\_weight & shadow\_mode\_review & canary\_rollout & v1.0.0-shadow \\
v1.2.0-prod & parameter\_update & hysteresis\_threshold & post\_incident\_refinement & post\_incident\_review & v1.1.0-prod \\
\bottomrule
\end{tabular}}
\end{table}
\end{revision}

\section{Case Study: SMS Verification Resilience}
\label{sec:case-study}

The SMS verification workflow is representative of many third-party integrations: a high-volume user-facing operation, more than one provider, regional differences, and direct user impact when provider behavior degrades. The simplified flow was: the application requested an SMS verification message, the SMS service asked the decision layer for an eligible provider, the selected provider accepted the message, the service published an outcome event, and the metrics aggregator updated rolling provider-health signals used by subsequent decisions.

Before adaptive routing, the system already had a primary provider and a secondary provider, along with the conventional operational controls expected in production systems. The secondary integration was valuable, but ordinary provider selection was still largely governed by a primary/secondary posture and incident-time operational decisions rather than by continuously evaluated business-outcome telemetry. During a primary-provider degradation, engineers still needed to validate the incident, reason about eligibility, and determine whether and how to shift traffic. This created a gap between the start of degradation and context-aware traffic movement.

After the change, each send and verification outcome produced telemetry. The decision layer used cached rolling metrics to compare eligible providers. When the primary provider degraded in a region or operation window, the factor list could shift eligible traffic toward the secondary provider subject to traffic caps and policy constraints. When the primary recovered, traffic could return gradually instead of all at once.

The observed operational value was not merely faster failover. It was a change in the control model. Provider choice moved from a conventional primary/secondary posture with incident-time interpretation toward a governed, observable, policy-driven decision loop. Engineers could focus on validating the incident, observing routing behavior, communicating status, and adjusting policy only when the automated decision was not appropriate. In the observed deployment scenario, adaptive routing reduced dependence on incident-time operator judgment and avoided widespread user-visible verification disruption during provider degradation windows. We state this conservatively because exact production telemetry cannot be disclosed publicly.

\subsection{Observed operational evidence}
\label{sec:observed-evidence}

\begin{revision}
This report uses permission-cleared, de-identified operational evidence extracts to support the industrial experience described here. The extracts use provider aliases, normalized indices, hashed request identifiers, and region groups. They are not raw production logs and should be interpreted as representative evidence supporting the experience report. Table~\ref{tab:before-after-evidence} summarizes four 15-minute windows: conventional baseline, shadow mode, adaptive routing under stable conditions, and adaptive routing during an active provider degradation.

\begin{table}[htbp]
\centering
\caption{Representative before/after operational windows from the de-identified evidence extract.}
\label{tab:before-after-evidence}
\scriptsize
\begin{tabular}{llrrrrrr}
\toprule
Window & Mode & Completion & Index & p95 ms & Retries & Fallbacks & Traffic A/B \\
\midrule
T-30d & conventional\_failover & 91.99\% & 100.0 & 380 & 370 & 112 & 100.0\%/0.0\% \\
T-15d & shadow\_mode & 92.10\% & 100.1 & 375 & 370 & 108 & 100.0\%/0.0\% \\
T+10d & adaptive\_routing & 95.81\% & 104.1 & 290 & 70 & 15 & 68.5\%/22.0\% \\
T+14d\_inc & adaptive\_routing & 92.00\% & 100.0 & 340 & 270 & 85 & 15.0\%/75.0\% \\
\bottomrule
\end{tabular}
\end{table}

Relative to the conventional baseline window, the stable adaptive-routing window shows a completion-rate index increase from 100.0 to 104.1, p95 latency reduction from 380 ms to 290 ms (23.7\% lower), retries reduced from 370 to 70 (81.1\% lower), and fallback invocations reduced from 112 to 15 (86.6\% lower). During the representative degradation window, the router reduced primary-provider share to 15.0\% and shifted 75.0\% to the secondary provider, while maintaining a completion-rate index of 100.0. These figures are not generalized beyond the anonymized windows; they provide concrete evidence for the type of operational improvement observed.

\begin{figure}[htbp]
  \centering
  \includegraphics[width=0.92\linewidth]{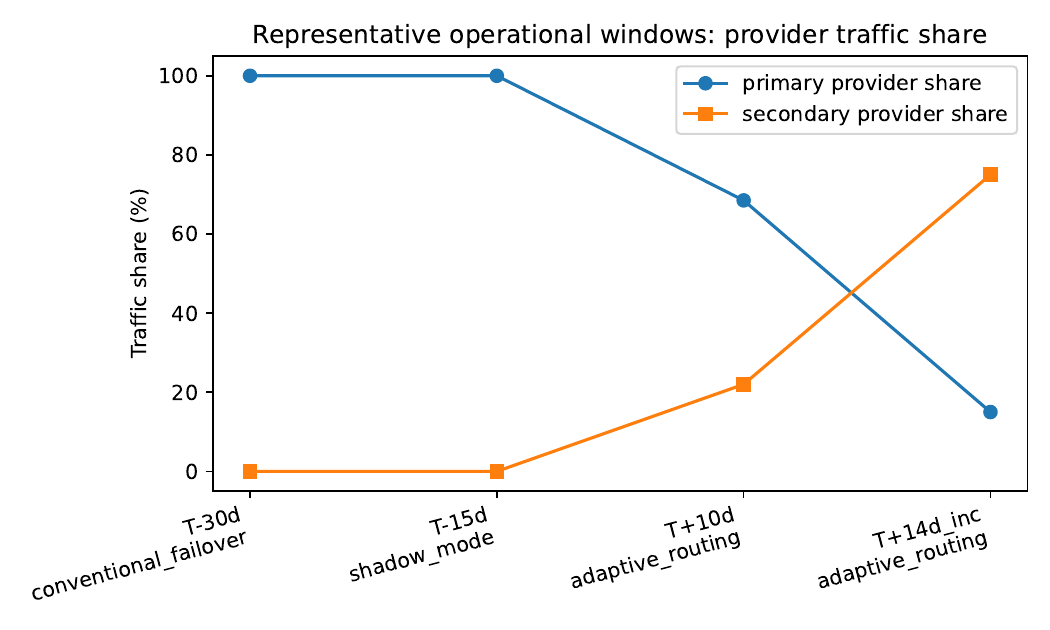}
  \caption{Representative operational windows showing how adaptive routing changed provider traffic share while preserving business completion signals.}
  \label{fig:traffic-share-windows}
\end{figure}

\subsection{Representative incident timeline}
\label{sec:incident-timeline}

Table~\ref{tab:incident-timeline} shows a relative incident timeline for a provider-degradation event. The event begins with elevated p95 latency and gateway timeouts. The router initially decays the provider score, then circuit state and traffic caps limit further movement. Probe traffic later detects recovery, a half-open state validates health, and failback completes after cooldown.

\begin{table}[htbp]
\centering
\caption{Representative anonymized incident timeline.}
\label{tab:incident-timeline}
\scriptsize
\begin{tabular}{llrrrrl}
\toprule
Time & Event & Comp. index & Latency index & Primary share & Action \\
\midrule
T-30m & stable\_baseline & 104.1 & 89.5 & 68.5\% & none \\
T0 & degradation\_detected & 88.4 & 165.0 & 45.0\% & score\_decay\_shift \\
T+5m & traffic\_shift\_capped & 99.2 & 94.0 & 15.0\% & circuit\_open\_shift\_to\_secondary \\
T+15m & recovery\_detected & 103.5 & 91.0 & 25.0\% & probe\_exploration\_increase \\
T+30m & traffic\_restored & 104.2 & 89.0 & 68.0\% & full\_failback\_completed \\
\bottomrule
\end{tabular}
\end{table}

\begin{figure}[htbp]
  \centering
  \includegraphics[width=0.92\linewidth]{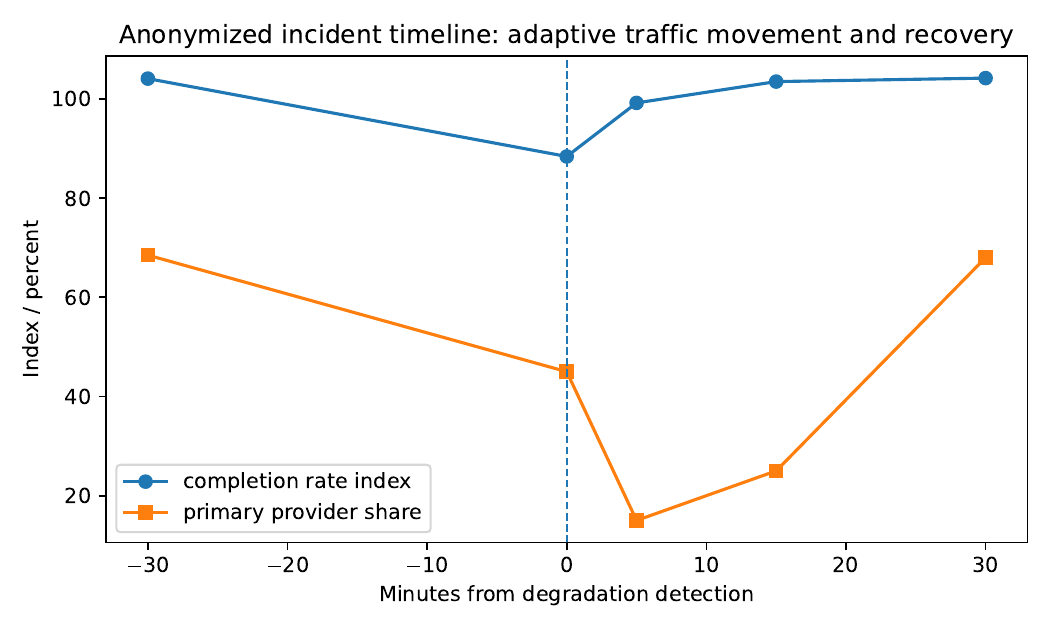}
  \caption{Incident timeline from the anonymized evidence extract. Completion-rate index recovered as traffic moved away from the degraded provider and later returned after probe-confirmed recovery.}
  \label{fig:incident-timeline}
\end{figure}
\end{revision}

\section{Challenges Encountered and Mitigations}
\label{sec:challenges}

The implementation exposed several challenges that are easy to underestimate. Table~\ref{tab:challenges} summarizes the most important ones and the mitigations used or recommended.

\begin{longtable}{>{\raggedright\arraybackslash}p{0.20\linewidth}>{\raggedright\arraybackslash}p{0.34\linewidth}>{\raggedright\arraybackslash}p{0.36\linewidth}}
\caption{Challenges and practitioner implications.}\label{tab:challenges}\\
\toprule
Challenge & Why it mattered & Mitigation and practitioner implication \\
\midrule
\endfirsthead
\toprule
Challenge & Why it mattered & Mitigation and practitioner implication \\
\midrule
\endhead
Stale metrics & Near-real-time telemetry can lag because of stream delays, backpressure, cache failures, partition rebalances, or clock skew. A stale healthy metric can keep traffic on a degraded provider. & Store metric timestamps, enforce expiry, penalize stale samples, and define safe defaults. Treat metric freshness as a gate or penalty, not an afterthought. \\
Small sample sizes & A provider with a perfect success rate over a few requests may be less trustworthy than a provider with a slightly lower rate over thousands of requests. & Include sample-size confidence and minimum sample thresholds. Use shadow mode to identify noisy windows before automation. \\
Circuit-breaker flapping & A provider can appear recovered under probe traffic, receive more traffic, and fail again. & Use half-open states, cooldown windows, traffic caps, and gradual recovery. Avoid binary full failback. \\
Regional heterogeneity & Provider health can differ by country, carrier, or operation type. A provider-wide average can hide regional degradation. & Aggregate metrics by provider, operation, and region where volume supports it. Fall back to coarser windows only when confidence is low. \\
Cost and reliability tension & The cheapest provider may create retries, user abandonment, or support load if completion is lower. & Treat cost as a score only after eligibility and reliability requirements are satisfied. Use different factor lists for critical and non-critical operations. \\
Policy ownership & Routing policy can involve SRE, engineering, product, compliance, finance, and vendor management. Unclear ownership can make changes slow or unsafe. & Define who owns gates, weights, override procedures, and review requirements. Version and audit policy changes. \\
Explainability & Operators may distrust automation if they cannot see why traffic moved. & Record reason codes, policy versions, gate outcomes, scores, and traffic-share changes. Put decision logs in incident dashboards. \\
Probe traffic & Routing away from a provider reduces information about whether it has recovered. & Allocate controlled probe traffic when policy allows. Separate recovery detection from full traffic restoration. \\
Fallback when all providers are ineligible & A strict gate model can exclude every provider during compound failures. & Define explicit fallback policies such as queue, retry later, degraded user messaging, or operator override. \\
Configuration mistakes & A bad factor list can move traffic incorrectly even if the code is correct. & Schema-validate configurations, test them in staging, use shadow mode, require review for high-impact changes, and provide rollback. \\
\bottomrule
\end{longtable}

These challenges also influenced the writing of this paper. We avoid presenting adaptive routing as a universal solution. It is useful when a critical operation has multiple eligible providers, enough traffic to generate meaningful telemetry, and an operating model capable of governing runtime policy. It can be overkill or risky when those conditions are absent.

\section{Supporting Evaluation: Synthetic Replay of Operational Scenarios}
\label{sec:evaluation}

The industrial case cannot disclose production logs, vendor names, internal thresholds, or exact incident data. To make the evaluation logic inspectable, we provide a deterministic synthetic artifact. The artifact is not intended to prove production impact. It is intended to help readers understand how the routing policy behaves under representative degraded-provider scenarios. The evaluation follows the same practical motivation as chaos-engineering and resilience-testing work: exercising behavior under injected degradation and turbulent operating conditions rather than relying only on nominal-path evidence \cite{owotogbe2026chaos}.

\subsection{Policies compared}

We compare four routing policies:

\begin{enumerate}[leftmargin=*]
  \item \textbf{Static primary:} all traffic uses the configured primary provider until requests fail.
  \item \textbf{Conventional operational failover:} traffic moves after an assumed detection, diagnosis, and operational action delay. This baseline abstracts common production failover procedures; it is not a claim that the prior system lacked automation.
  \item \textbf{Circuit-breaker only:} local circuit state prevents calls to a provider after failure thresholds are crossed.
  \item \textbf{Factor-list routing:} gates and weighted scores use completion rate, latency, quota, incident penalty, and confidence signals, with traffic-shift controls.
\end{enumerate}

The simulated scenarios are complete provider outage, latency spike, regional failure, quota exhaustion, partial completion-rate degradation, and stale telemetry. The workload uses 36,000 synthetic requests per scenario and policy. The artifact includes the generator, input factor-list example, routing trace, aggregated metrics, and output figures.

\subsection{Evaluation results}

Table~\ref{tab:eval} summarizes the synthetic results. The table should be interpreted as supporting evidence for behavior under controlled assumptions, not as production performance data.

\begin{table}[htbp]
\centering
\caption{Synthetic evaluation summary. Success rate is shown as percentage of synthetic requests. Lower degraded minutes indicates faster movement away from the degraded provider.}
\label{tab:eval}
\scriptsize
\begin{tabular}{lrrrr}
\toprule
Scenario & Static primary & Conventional failover & Circuit breaker & Factor list \\
\midrule
Complete outage & 66.3\% / 20 min & 82.7\% / 10 min & 92.7\% / 4 min & 96.0\% / 4 min \\
Latency spike & 99.4\% / 20 min & 99.3\% / 14 min & 99.5\% / 20 min & 98.8\% / 2 min \\
Partial degradation & 94.8\% / 20 min & 97.1\% / 10 min & 98.5\% / 4 min & 98.5\% / 2 min \\
Quota exhaustion & 47.4\% / 32 min & 88.5\% / 7 min & 88.6\% / 7 min & 99.0\% / 0 min \\
Regional failure & 90.1\% / 20 min & 94.6\% / 10 min & 97.5\% / 4 min & 97.9\% / 3 min \\
Stale telemetry & 94.0\% / 17 min & 96.0\% / 10 min & 98.1\% / 4 min & 98.4\% / 2 min \\
\bottomrule
\end{tabular}
\end{table}

The results illustrate three points. First, the static-primary policy performs poorly in scenarios where the primary provider remains configured despite degradation. Second, conventional operational failover improves outcomes but depends on detection, diagnosis, and action delay when degradation is nuanced rather than binary. Third, the factor-list policy is most useful when the degradation is not just a binary failure, for example quota exhaustion, regional failure, or partial completion-rate degradation. The latency-spike scenario also shows an important caution: a factor-list policy can trade a small amount of success-rate performance for faster movement away from degraded latency if weights overemphasize latency. That behavior is not automatically good or bad; it is a policy decision that must reflect the operation's service objective.

\begin{figure}[htbp]
  \centering
  \includegraphics[width=0.92\linewidth]{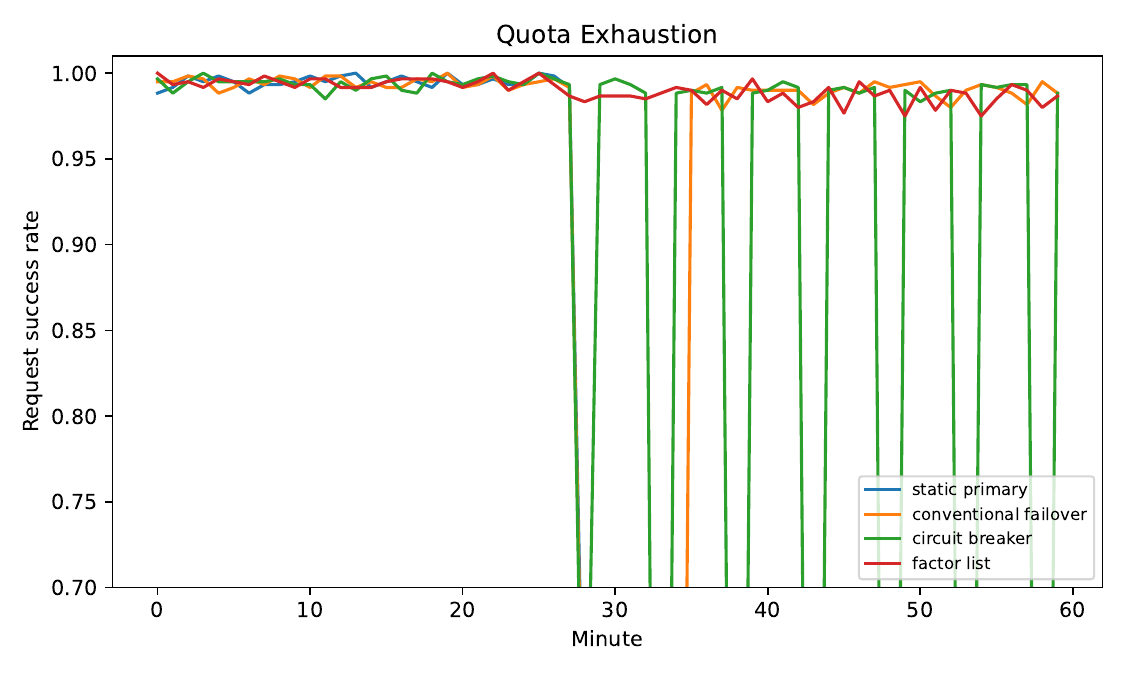}
  \caption{Synthetic quota-exhaustion scenario. Factor-list routing can react before transport-level failures dominate because quota availability is treated as an eligibility gate.}
  \label{fig:quota}
\end{figure}

\begin{figure}[htbp]
  \centering
  \includegraphics[width=0.92\linewidth]{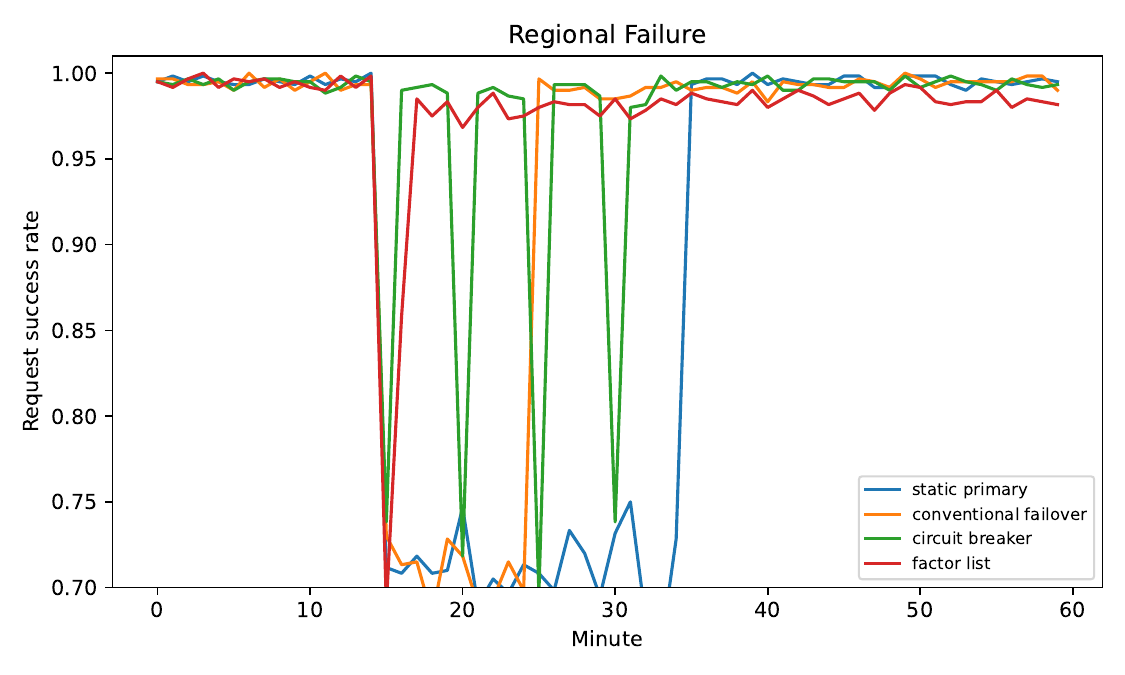}
  \caption{Synthetic regional-failure scenario. Region-aware scoring and gates avoid treating provider health as a single global value.}
  \label{fig:regional}
\end{figure}

\begin{revision}
\subsection{Sensitivity analysis}
\label{sec:sensitivity}

The supplementary material includes a sensitivity table that varies completion-rate weight, latency weight, traffic cap, and metric staleness. Table~\ref{tab:sensitivity} summarizes the observed behavior in the replay output. The results indicate that the policy is robust to moderate weight shifts but sensitive to traffic caps and telemetry delay. Reducing the traffic cap slows movement away from a degraded provider, while increasing metric staleness increases oscillation and degraded-provider traffic share.

\begin{table}[htbp]
\centering
\caption{Sensitivity analysis for selected routing parameters.}
\label{tab:sensitivity}
\scriptsize
\resizebox{\linewidth}{!}{%
\begin{tabular}{llrrrrrr}
\toprule
Scenario & Parameter & Completion idx & Failed idx & p95 idx & Switches & Osc. & Shift-away min \\
\midrule
scen\_01\_baseline & none & 104.1 & 82.0 & 89.5 & 12 & 0 & 4.5 \\
scen\_02\_comp\_heavy & weight\_completion\_rate & 104.4 & 79.5 & 93.2 & 14 & 1 & 3.8 \\
scen\_03\_lat\_heavy & weight\_latency & 103.8 & 84.1 & 85.1 & 18 & 2 & 3.2 \\
scen\_04\_low\_cap & traffic\_cap\_percent & 101.8 & 94.0 & 96.0 & 12 & 0 & 12.0 \\
scen\_05\_stale\_delay & metric\_staleness\_seconds & 102.5 & 91.2 & 94.5 & 8 & 3 & 8.5 \\
\bottomrule
\end{tabular}}
\end{table}

\begin{figure}[htbp]
  \centering
  \includegraphics[width=0.90\linewidth]{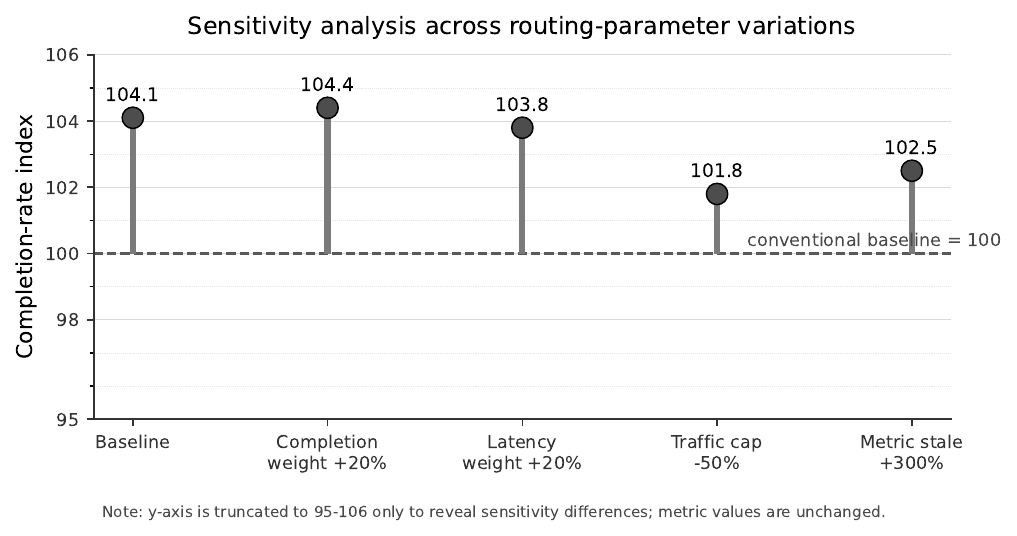}
  \caption{Sensitivity-analysis output for completion-rate index under representative routing-parameter variations.}
  \label{fig:sensitivity}
\end{figure}
\end{revision}

\subsection{What the evaluation does and does not show}

The evaluation helps readers inspect the routing logic under repeatable conditions. It does not replace a controlled production experiment, and it does not claim that the exact percentages generalize to other organizations. The key practical lesson is that adaptive routing must be tested against failure modes that resemble operations: quota exhaustion, stale telemetry, partial degradation, and regional failure are often more relevant than a clean provider-down event.

\section{Lessons Learned and Practitioner Guidance}
\label{sec:lessons}

The experience led to several lessons that may transfer to other critical third-party integrations.

\subsection{What we would do differently}

Several improvements would have reduced adoption risk or accelerated learning.

\begin{itemize}[leftmargin=*]
  \item \textbf{Define outcome metrics earlier.} Provider-call success, carrier delivery, and user verification completion are related but not identical. Defining the primary routing objective earlier would have reduced debate during shadow-mode validation.
  \item \textbf{Make decision logs first-class from the start.} Decision logs became essential for trust. They should be treated as product requirements for the router, not as observability work added after the routing policy is implemented.
  \item \textbf{Separate policy ownership before tuning weights.} Weight tuning is less useful if teams have not agreed who owns regional gates, quota limits, cost objectives, and incident overrides.
  \item \textbf{Test recovery as carefully as failover.} Moving away from a degraded provider is only half the problem. Gradual recovery, probe budgets, and cooldown windows should be validated before production enablement.
  \item \textbf{Prepare a negative case.} The rollout should explicitly identify operations where adaptive routing is not justified. This prevents the pattern from becoming an overgeneralized platform mandate.
\end{itemize}

\subsection{Lessons learned}

\begin{enumerate}[leftmargin=*]
  \item \textbf{A backup provider and a conventional failover plan are not sufficient by themselves.} Resilience requires a safe and timely decision about when to use each eligible provider, how much traffic to move, and when to return.
  \item \textbf{Business outcomes should influence routing.} API response codes are useful, but user-flow completion, downstream delivery, and operation-specific success are often better routing signals.
  \item \textbf{Gates and scores should be separated.} Compliance, region support, quota, credential validity, and circuit state should exclude a provider before optimization occurs.
  \item \textbf{Automation needs explainability.} Operators are more likely to trust adaptive routing when dashboards show the reason code, policy version, gate outcomes, score components, and traffic-share change.
  \item \textbf{Shadow mode reduces adoption risk.} Computing what the router would have done without moving traffic is a practical way to test policy logic and build trust.
  \item \textbf{Dampening is part of correctness.} Hysteresis, traffic caps, probe traffic, and cooldowns are not optional add-ons; they prevent unstable feedback loops.
  \item \textbf{Ownership must be explicit.} A routing policy is a socio-technical artifact. Engineering, SRE, compliance, finance, and product concerns can all affect the factor list.
  \item \textbf{Do not automate before observability.} If teams cannot explain current provider behavior, automating provider selection will likely create a harder-to-debug system.
\end{enumerate}

\subsection{Practitioner checklist}

Table~\ref{tab:checklist} translates these lessons into a checklist for teams considering adaptive provider routing.

\begin{table}[htbp]
\centering
\caption{Practitioner checklist for deciding whether and how to adopt adaptive API routing.}
\label{tab:checklist}
\small
\begin{tabularx}{\linewidth}{L{0.30\linewidth}Y}
\toprule
Question & Practical guidance \\
\midrule
Is the operation critical? & Prioritize flows where provider failure blocks login, account recovery, payment authorization, safety checks, or other high-impact journeys. \\
Are there truly eligible alternatives? & Do not build adaptive routing if alternate providers lack coverage, compliance approval, quota, or equivalent behavior. \\
Is telemetry volume sufficient? & Dynamic decisions need enough samples by operation and region. Low-volume flows may need conservative windows or operator-controlled routing. \\
Are business outcomes observable? & Instrument completion or user-flow outcome where possible, not only transport-level status. \\
Can operators override automation? & Provide kill switches, provider pinning, rollback, and emergency policy controls. \\
Can decisions be explained? & Log reason codes, policy versions, gate outcomes, scores, and selected provider. \\
Is configuration governed? & Schema-validate factor lists and require review for changes that can materially shift traffic. \\
Has shadow mode been run? & Compare router recommendations with incidents and normal traffic before enabling production routing. \\
Are recovery rules safe? & Use half-open circuits, probe traffic, cooldowns, and traffic caps to avoid flapping. \\
Is the architecture proportionate? & For low-risk or low-volume operations, a conventional primary/secondary failover model with tested runbooks may be simpler and safer. \\
\bottomrule
\end{tabularx}
\end{table}

\section{Transferability, Limitations, and Threats to Validity}
\label{sec:transferability}

\subsection{Transferability conditions}

The approach is most likely to transfer when the following conditions hold:

\begin{itemize}[leftmargin=*]
  \item The operation is user-facing or business-critical.
  \item More than one provider is technically and contractually eligible.
  \item Provider behavior varies by region, operation type, quota state, or incident window.
  \item The platform has enough traffic to compute meaningful rolling metrics.
  \item Teams can instrument business outcomes and provider-call outcomes.
  \item There is organizational ownership for policy review, incident override, and compliance gates.
\end{itemize}

The approach is less appropriate when there is no true alternate provider, traffic volume is too low to support confidence, failure impact is low, compliance rules cannot be encoded safely, or the organization lacks operational maturity to govern runtime routing policy. In those cases, a simpler design with timeouts, retries, circuit breakers, dashboards, static priorities, and tested incident runbooks may be the better engineering choice.

\subsection{Transferability by operation type}

Table~\ref{tab:transferability} gives a practitioner-oriented view of where the pattern is likely to transfer and what must change.

\begin{table}[htbp]
\centering
\caption{Transferability considerations for other third-party API integrations.}
\label{tab:transferability}
\small
\begin{tabularx}{\linewidth}{L{0.22\linewidth}YY}
\toprule
Operation type & Transferability assessment & Additional controls likely needed \\
\midrule
SMS or email verification & Strong fit when completion and delivery signals exist and providers have overlapping regional coverage. & Region and carrier gates, sample-size confidence, user-flow completion signals. \\
Payment authorization & Possible but higher risk because routing affects money movement, compliance, idempotency, and dispute handling. & Strict compliance gates, reconciliation, idempotency controls, audit evidence, conservative traffic caps. \\
Fraud or risk checks & Possible when alternate providers produce comparable risk signals and false-positive costs are understood. & Model/version governance, fairness and risk review, outcome calibration, human-review escalation. \\
Observability or notification APIs & Often a moderate fit; user impact may be indirect, and simpler buffering may be enough. & Queueing, backpressure, data-loss policies, lower-cost recovery paths. \\
Fulfillment or tax integrations & Context-dependent; provider substitutability and contractual constraints often dominate. & Contract and jurisdiction gates, retry/compensation logic, domain-specific reconciliation. \\
\bottomrule
\end{tabularx}
\end{table}

\subsection{Threats to validity}

\textbf{Internal validity.} The synthetic evaluation depends on modeled provider behavior, metric windows, telemetry delays, and policy weights. Different assumptions could produce different results. We mitigate this by making the synthetic artifact deterministic and inspectable.

\textbf{External validity.} The industrial case concerns SMS verification in a marketplace environment. The same architecture may not transfer directly to payments, fraud checks, identity verification, observability, or fulfillment integrations without changing gates, metrics, risk controls, and compliance constraints.

\textbf{Construct validity.} Completion rate is an important signal but does not capture every user outcome. For example, successful SMS delivery may not imply successful login if the user abandons the flow. Conversely, a failed provider request may be retried successfully. Teams should define operation-specific success carefully.

\textbf{Operational validity.} Adaptive routing can create new failure modes, including stale-metric decisions, circuit flapping, cost spikes, quota exhaustion, or unstable traffic movement. The architecture should therefore be introduced incrementally and paired with dashboards, kill switches, and policy governance.

\section{Data Availability and Reproducibility Artifact}

\begin{revision}
The supplementary artifact contains two kinds of evidence. First, it contains permission-cleared, de-identified operational evidence extracts used to support the industrial experience report. These files include before/after operational windows, routing-decision traces, provider-call outcomes, business-completion outcomes, shadow-mode validation records, policy-version history, incident timeline records, probe/exploration traffic records, circuit/quota gate state, score-normalization snapshots, sensitivity-analysis outputs, contribution-boundary mapping, and a data dictionary. Second, it contains the synthetic replay artifact used to make routing behavior inspectable without exposing production logs.

The operational evidence files use provider aliases, hashed identifiers, normalized indices, region groups, and relative incident time. Raw production logs, exact vendor names, user identifiers, phone numbers, message payloads, internal system names, contractual pricing, proprietary thresholds, and internal infrastructure details remain unavailable due to confidentiality, privacy, contractual, and commercial restrictions.
\end{revision}

\section{Conclusion}

This industrial experience suggests that third-party API resilience is not achieved by integrating an alternate provider and operating a conventional failover plan alone. Those mechanisms remain necessary, but the practical challenge is deciding when a provider is eligible, when it is healthy enough, how quickly traffic should move, and how operators can understand and override the decision during incidents.

Pluggable factor lists provided a useful control-plane abstraction in the reported setting. Gates encoded non-negotiable constraints, scores encoded preferences among eligible providers, telemetry converted outcomes into routing signals, and decision logs made automated choices explainable. The most important lesson is not that every integration should become adaptive. Rather, critical integrations with multiple eligible providers, sufficient telemetry, and strong operational ownership can benefit from moving routine provider selection beyond a static primary/secondary posture and into a governed, observable, runtime decision process.

\section*{CRediT authorship contribution statement}

Nataraj Agaram Sundar: Conceptualization, Methodology, Software, Formal analysis, Investigation, Visualization, Writing - original draft, Writing - review and editing, Project administration. Tejas Morabia: Conceptualization, Methodology, Investigation, Validation, Writing - review and editing, Supervision, Project administration.

\section*{Declaration of competing interest}

The authors declare the following potential competing interest: this manuscript is based on permission-cleared and anonymized industry experience related to high-scale service integrations at eBay Inc. Nataraj Agaram Sundar and Tejas Morabia are employed by eBay Inc. The manuscript does not disclose vendor names, customer identifiers, internal system names, confidential telemetry, or proprietary operational thresholds. No external vendor, platform, or third-party service provider funded the work or participated in manuscript preparation. The authors declare no other known competing financial interests or personal relationships that could have appeared to influence the work reported in this paper.

\section*{Funding}

This research did not receive any specific grant from funding agencies in the public, commercial, or not-for-profit sectors.

\section*{Declaration of generative AI and AI-assisted technologies in the manuscript preparation process}

During the preparation of this work, the authors used OpenAI ChatGPT to assist with reviews. After using this tool, the authors reviewed and edited the content as needed and take full responsibility for the content of the submitted manuscript.

\section*{Acknowledgements}

The authors thank the engineering and operations teams whose practical experience with high-scale service integrations informed the architecture described in this paper. All vendor names, internal system names, and organization-specific operational details have been anonymized.

\end{document}